\documentclass[sigplan]{acmart}
\renewcommand\footnotetextcopyrightpermission[1]{}
\AtBeginDocument{%
  }

\setcopyright{acmlicensed}
\copyrightyear{2018}
\acmYear{2018}
\acmDOI{XXXXXXX.XXXXXXX}
\acmISBN{978-1-4503-XXXX-X/2018/06}

\usepackage{lzj-style}
\usepackage{balance}
\usepackage{soul}
\newcommand{\NAME}[1]{{\texttt{Bullet}}}

\usepackage{etoolbox}  
\SetAlCapSkip{2em}       
\SetAlgoSkip{smallskip}  

\begin{document}


\title{Boosting LLM Serving through Spatial-Temporal GPU Resource Sharing}

\settopmatter{authorsperrow=3}
\author{Zejia Lin}
\orcid{0000-0002-7205-4062}
\affiliation{%
  \institution{Sun Yat-sen University}
  \city{Guangzhou}
  \country{China}
}
\email{linzj39@mail2.sysu.edu.cn}

\author{Hongxin Xu}
\affiliation{%
  \institution{Sun Yat-sen University}
  \city{Guangzhou}
  \country{China}
}
\email{xuhx56@mail2.sysu.edu.cn}

\author{Guanyi Chen}
\affiliation{%
  \institution{Sun Yat-sen University}
  \city{Guangzhou}
  \country{China}
}
\email{chengy259@mail2.sysu.edu.cn}

\author{Zhiguang Chen}
\affiliation{%
  \institution{Sun Yat-sen University}
  \city{Guangzhou}
  \country{China}
}
\email{chenzhg29@mail.sysu.edu.cn}

\author{Yutong Lu}
\orcid{0000-0001-5315-3375}
\affiliation{%
  \institution{Sun Yat-sen University}
  \city{Guangzhou}
  \country{China}
}
\email{luyutong@mail.sysu.edu.cn}

\author{Xianwei Zhang}
\authornote{Corresponding author.}
\orcid{0000-0003-3507-4299}
\affiliation{%
  \institution{Sun Yat-sen University}
  \city{Guangzhou}
  \country{China}
}
\email{zhangxw79@mail.sysu.edu.cn}

\renewcommand{\shortauthors}{}

\begin{abstract}
Modern large language model (LLM) serving systems confront inefficient GPU utilization due to the fundamental mismatch between compute-intensive prefill phase and memory-bound decode phase. 
While current practices attempt to address this by organizing these phases into hybrid batches, such solutions create an inefficient tradeoff that sacrifices either throughput or latency,  leaving substantial GPU resources underutilized.
For this, we identify two key root causes: 1) the prefill phase suffers from suboptimal compute utilization due to wave quantization and attention bottlenecks, and
2) hybrid batching disproportionately prioritizes latency over throughput, wasting both compute resources and memory bandwidth.
To mitigate the issues, we present \NAME{}, a novel spatial-temporal orchestration system that eliminates these inefficiencies through fine-grained phase coordination.
\NAME{} enables concurrent execution of prefill and decode requests, while dynamically provisioning GPU resources based on real-time performance modeling.
By integrating SLO-aware scheduling and adaptive resource allocation, \NAME{} maximizes GPU utilization without compromising latency targets.
Experimental evaluations on real-world workloads demonstrate that \NAME{} delivers 1.26$\times$ average throughput gains (up to 1.55$\times$) 
over state-of-the-arts, while consistently meeting latency constraints.
\end{abstract}



\maketitle
\section{Introduction}\label{sec:intro}

GPUs have become the predominant computing platform for large language model (LLM) services \cite{chatgpt,dsv3,llama}, powering a wide range of applications with varying computational and latency demands \cite{metagpt,llamaindex}.
As these applications continue to grow in scale and complexity, maximizing GPU utilization has become crucial for elevating service quality \cite{nanoflow,parrot}.
In response, a plethora of systems have been developed to optimize different aspects of LLM serving, such as kernel-level performance \cite{vllm,flashattn,podattn}, scheduling strategies \cite{sarathi,distserve,splitwise}, and parallelization techniques \cite{loongserve}.

\begin{figure}[!t]
    \centering
    \includegraphics[width=\linewidth]{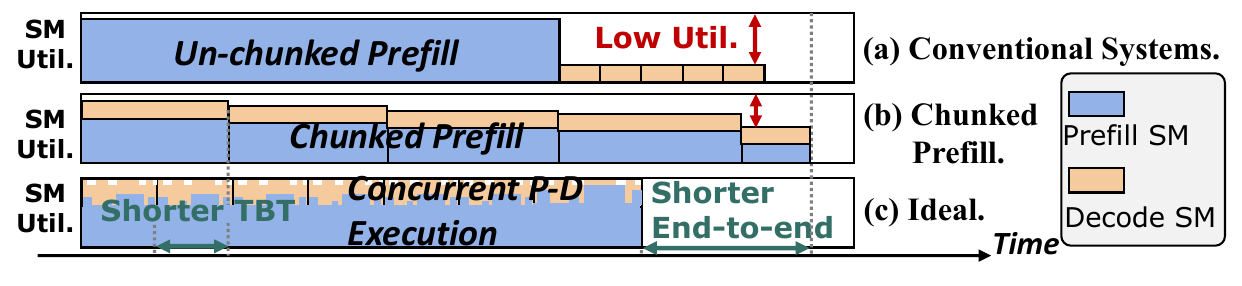}
    \vspace{-2em}
    \caption{Request-level computation of conventional systems (top), chunked prefill (middle) and ideal (bottom).}
    \label{fig:intro.compare}
    \vspace{-2em}
\end{figure}

However,  the divergent computational characteristics of LLM inference make high GPU resource utilization particularly challenging \cite{nanoflow}. 
In detail, the workflow consists of a computationally intensive \textit{prefill} phase that processes all inputs in parallel, succeeded by a memory-bound \textit{decode} phase that generates tokens sequentially.
These two contrasting phases lead to a structural imbalance in the utilization of computational resources and memory bandwidth.
The dynamic nature of workload patterns 
further exacerbates this imbalance, forcing GPUs to alternate between these complementary resource utilization states.
Moreover, the distinct characteristics of the two phases inherently introduce a {\bf throughput-latency tradeoff} \cite{sarathi}. 
For example, conventional systems \cite{vllm} (Figure \ref{fig:intro.compare}a) prioritize the prefill, which thus increases the size of subsequent decode batches. 
While this improves throughput, it also penalizes latency, since prefill operations monopolize GPU resources.
Therefore, an ideal system necessitates fine-grained control to balance throughput and latency without elevated GPU utilization, and is non-trivial to achieve.

Prior attempts have sought to address phase imbalance either by isolating prefill and decode across different GPUs, or by coordinating them on the same device through \textit{spatial} multiplexing.
Prefill-decode disaggregation \cite{distserve,splitwise,dsv3} represents the first approach, physically separating the two phases across dedicated GPU groups.
However, such systems require careful tuning of GPU allocations for each phase, tailored to specific workload patterns \cite{dsv3}, and struggle to reconfigure quickly under fluctuating request loads \cite{distserve}.
Furthermore, they impose heavy demands on high-bandwidth interconnects to support frequent state migration \cite{mooncake}, significantly restricting the deployment scenarios.
In practice, chunked prefill \cite{sarathi} (Figure \ref{fig:intro.compare}b) has been pervasively adopted in production systems \cite{vllm,sglang,trtllm,mii}.
This method leverages a fixed token budget to combine prefill and decode requests into \textit{hybrid batches}, with longer sequences being partitioned into chunks to fit within capacity.
While tunable chunk sizes provide a degree of latency control, the approach inevitably compromises throughput \cite{sarathi,podattn}. 
Smaller chunk sizes often underutilize GPU capacity, while larger ones negate latency improvements, reflecting the inherent trade-off in balancing efficiency and responsiveness.

Although throughput-latency tradeoff in LLM serving has been extensively studied \cite{sarathi,distserve}, we identify a critical limitation that, existing approaches often underutilize GPUs, resulting in suboptimal balance between latency and hardware efficiency.
First, the attention exhibits relatively low compute utilization compared to linear layers (\cref{sec:bg.principle}) despite optimized implementations \cite{flashattn}.
Second, the throughput-latency tradeoff in chunked prefill exhibits sub-linear scaling with chunk sizes that prolongs execution time for successive chunks (\cref{sec:bg.biased}).
Third, while spatial sharing has the potential to exploit the natural complementarity between compute-intensive prefill and memory-bound decode phases through concurrent execution (Figure \ref{fig:intro.compare}c), effective orchestration is required.
Without careful coordination, contention and latency violations remain significant risks.
These observations underscore the need for dynamic coordination to fully exploit GPU performance.
To address GPU under-utilization and balance the trade-off between throughput and latency, we propose \NAME{}, an LLM serving system that saturates GPU resources through \ul{\bf spatial-temporal orchestration with fine-grained resource provisioning}.
\NAME{} proactively monitors request progress and dynamically adjusts resource provision to sustain high utilization while satisfying latency requirements.
The system  achieves such efficient execution with four key components: 1) a performance estimator (\cref{sec:ds.estim}) with an accurate analytical model of low profile and runtime overhead; 2) an SLO-aware task scheduler (\cref{sec:ds.sched}) that dynamically adjusts prefill and decode requests to balance throughput and latency; 3) a computational resource manager (\cref{sec:ds.smpart}) that offers lightning yet precise resource configuration; 4) a concurrent execution engine (\cref{sec:ds.engine}) that enables asynchronous CPU control flow and GPU execution.

\NAME{} achieves optimal resource provisioning via an SM-scaling roofline model that captures contention between co-executed prefill and decode kernels with varying computational resource partitions.
Calibrated with minimal samples, the model is capable of continuous refinement with online data.
Layer-wise prefill scheduling monitors request progress and SLO compliance, reconfiguring resources through lightweight compute-unit masking \cite{libsmctrl}.
With microsecond-level operational overhead, \NAME{} maintains both precise control and high efficiency throughout execution.
\NAME{} is available at \url{https://github.com/zejia-lin/Bullet}.

In summary, the paper makes the following contributions: \vspace*{-1em}
\begin{itemize}
    \item We identify the inefficiencies that hinder GPU utilization in existing LLM serving systems while navigating throughput-latency tradeoffs, and systematically analyze opportunities for boosting utilization.
    \item We establish accurate modeling for spatial-temporal shared phases, and introduce fine-grained control mechanisms for latency-aware resource provisioning.
    \item We design and implement an LLM serving system to effectively integrate the proposed techniques into readily available frameworks.
    \item Experimental evaluations on real-world workloads show that \NAME{} outperforms state-of-the-arts in both latency and throughput while achieving significantly higher GPU utilization.
\end{itemize}

\section{Background and Motivation}

\subsection{LLM Computational Workflow}

Recent LLMs \cite{chatgpt,llama,qwen} are mainly built by stacking Transformer blocks \cite{alluneed}, with each containing four core components: QKV-projection layer, self-attention computation, Output-projection layer and multi-layer perception (MLP).
These components are primarily implemented through general matrix multiplications (GEMMs), with element-wise operations interspersed between layers.
Among them, the self-attention specifically operates on query, key, and value matrices produced by the QKV-projection. Its computational efficiency has been significantly enhanced through optimized kernel implementations like FlashAttention \cite{flashattn}.

The LLM inference pipeline operates through two distinct computational phases. 
During the initial prefill phase, the system processes all input tokens in parallel to generate the first output token, with the latency measured as time-to-first-token (TTFT). 
This compute-intensive stage performs full attention computations across the entire sequence while building the KV cache to store intermediate key-value states. 
Subsequently, the decode phase generates tokens sequentially, with each iteration consuming only the most recent output token to produce the next.
The average iteration latency is termed time-per-output-token (TPOT). 
Unlike prefill, this memory-bound phase primarily retrieves data from the KV cache and performs relatively lightweight computations for the current token's transformations. 

\subsection{GPU Utilization Principles}\label{sec:bg.principle}

\subsubsection{Execution Model and Theoretical Performance Bound.}\label{sec:bg.gpu}

Modern GPUs employ a hierarchical architecture with hundreds of streaming multiprocessors (SMs), each containing general-purpose cores and specialized matrix units like Tensor Cores \cite{h100}.
GPU's grid-block-thread programming model \cite{cuda} aligns with this architecture, where kernels are organized into grids of thread blocks (TBs) that each manage thousands of cooperating threads.
Upon kernel launch, it enters an asynchronous task queue (termed \textit{stream} in CUDA\cite{cuda}) for scheduling.
The hardware scheduler retrieves kernels from these queues and dispatches them across SMs, enabling the concurrent kernel execution (CKE) \cite{smk,cudasched,mostroom} from different streams when required resources are satisfied.

During execution, multiple TBs can reside on the same SM to share its registers, shared memory, and thread slots.
The TBs per SM can be obtained via hardware vendor's runtime APIs \cite{cuda,hipstream}.
The SM executes the \textit{warps} (32-thread groups) in successive \textit{wavefronts} to interleave different instructions and maximize hardware utilization.
However, if the number of TBs is not evenly divisible by the number of SMs, a workload imbalance situation called \textit{wave quantization} \cite{wavequant,asplos_forcast} occurs.
Some SMs finish early and remain idle while waiting for others to complete the tail wave.
Formally, given a kernel with $g$ TBs, $N$ SMs, and $b$ TBs per SM, the kernel demands $w=\lceil g / (b\cdot N)\rceil$ waves to complete.
In the final wave, the TBs are distributed unevenly due to Most-Room Policy \cite{mostroom,libsmctrl3}, resulting in only $tail =\left\lceil g/b - N\cdot(w - 1)\right\rceil$ SMs active.
Therefore, the corresponding ratio of idled SM cycles can be quantified as $idle = (N-tail)/(Nw)$.

GPU kernels typically use power-of-2 grid sizes to match data dimensions, but this conflicts with non-power-of-2 SM counts in GPUs.
For example, 108 for Nvidia A100 \cite{a100} and 132 for H100 \cite{h100}.
Therefore, wave quantization remains an open issue \cite{streamk,cusync} across diverse kernels.
This inefficiency is particularly pronounced in Transformer's self-attention \cite{alluneed,flashattn} and small-shaped GEMMs for short input sequences or small \textit{chunked prefill-sizes} (\cref{sec:bg.chunk_workflow}) \cite{sarathi}.
Additional under-utilization cause arises from memory-bound kernels like LLMs' decode phases and element-wise operators, which idle compute resources during frequent memory accesses.

\renewcommand{\arraystretch}{0.8}
\begin{table}[!b]
    \centering
    \vspace{-1em}
    \caption{Theoretical SM idle ratio (\%) caused by wave quantization effects, normalized to kernel/layer's execution time.}
    \vspace{-1em}
    \begin{tabular}{c|cccc|c}
    \toprule
    \textbf{Seq. Len.} & \textbf{QKV} & \textbf{Attn} & \textbf{O} & \textbf{MLP} & \textbf{Layer's Total} \\
    \midrule
    1024 & 11.1 & 21.0 & 40.7 & 13.0 & 19.4 \\
    2048 & 11.1 & 5.2 & 21.0 & 7.6 & 10.4 \\
    4096 & 11.1 & 5.2 & 5.2 & 7.6 & 9.1 \\
    16384 & 1.9 & 0.2 & 0.2 & 0.4 & 0.5 \\
    \bottomrule
    \end{tabular}
    \label{tab:bg.wave_quant}
\end{table}

\subsubsection{LLM Kernel Characteristics}\label{sec:bg.kernel}

\begin{figure}[!t]
    \centering 
    \includegraphics[width=\linewidth]{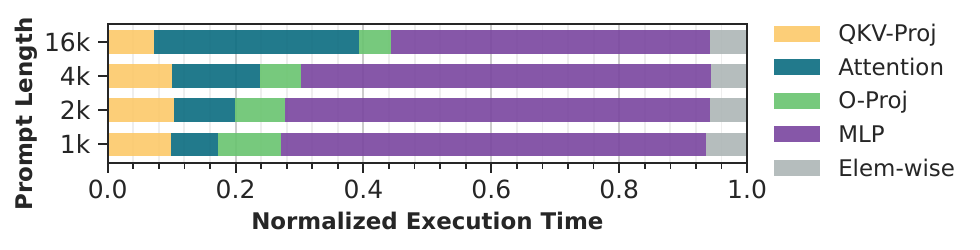}
    \begin{subfigure}[t]{\linewidth}
        \vspace{-2em}
        \caption{Execution time. (CPU overhead excluded)}
        \label{fig:bg.timebreak}
    \end{subfigure}
    \includegraphics[width=\linewidth]{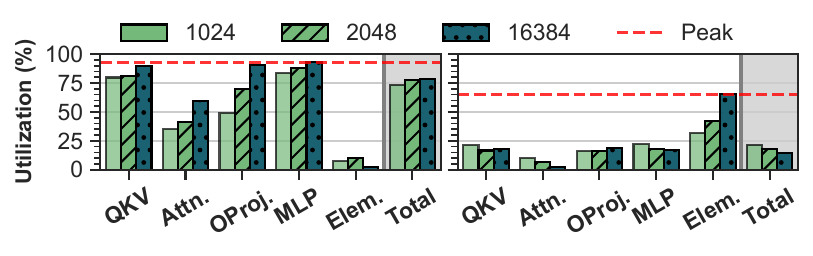}
    \begin{subfigure}[t]{.44\linewidth}
        \vspace{-2em}
        \caption{Compute utilization.}
    \end{subfigure}
    \begin{subfigure}[t]{.55\linewidth}
        \vspace{-2em}
        \caption{Memory bandwidth utilization.}
        \vspace*{-1em}
    \end{subfigure}
    \vspace*{-2em}
    \caption{Breakdown of execution time and hardware utilization in the prefill phase of Llama-3.1-8B model on Nvidia A100 GPU. The aggregate utilization per layer remains below peak sustainable capacity (red line).}
    \label{fig:bg.kernel_profile}
    \vspace{-1.5em}
\end{figure}

We quantify LLM serving efficiency by analyzing
execution time (Figure \ref{fig:bg.timebreak}), hardware utilization (Figure \ref{fig:bg.kernel_profile}b,c), and the theoretical waste caused by wave quantization (Table \ref{tab:bg.wave_quant}).
While MLP operations achieve up to 92\% compute utilization, complete Transformer layers sustain only 70\%-76\% due to compounding inefficiencies. 
For shorter sequences, severe wave quantization in GEMMs creates substantial underutilization, as evidenced by the O-proj's measured 49\% and 70\% utilization, respectively.
This result closely agrees with the theoretical bound of 59\% and 79\% ($100-idle\%$ in Table \ref{tab:bg.wave_quant}).
The attention kernels, even with optimized implementations \cite{flashattn}, exhibit lower utilization than GEMM and account for 34\% of the runtime on long sequences.
Together, the effects of wave quantization and attention bottleneck create persistent performance gaps between theoretical peak and achieved throughput. 
These gaps remain substantial regardless of sequence length, inherently constraining overall system efficiency.

\subsubsection{GPU Resource Provisioning and Sharing.}\label{sec:bg.gpushare}

\begin{figure}[!b]
    \centering
    \vspace{-1em}
    \includegraphics[width=\linewidth]{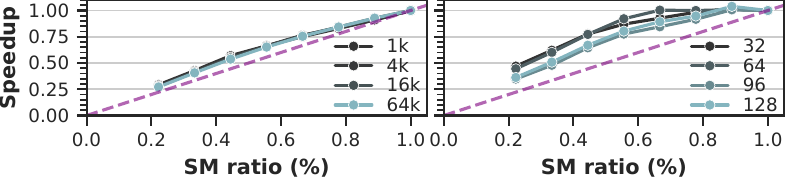}
    \begin{subfigure}[t]{.44\linewidth}
        \vspace{-1.5em}
        \caption{Prefill phase.}
    \end{subfigure}
    \begin{subfigure}[t]{.55\linewidth}
        \vspace{-1.5em}
        \caption{Decode phase. (CL: 2048)}
        \vspace*{-1em}
    \end{subfigure}
    \vspace{-2em}
    \caption{Speedup of using partial SMs normalized to using full GPU. (Purple dashed line: linear scale.)}
    \label{fig:bg.sm_speedup}
    
\end{figure}

Naturally, compute- and memory-bound kernels are suitable to co-execute on GPUs, saturating both compute and bandwidth resources \cite{reef,orion,gpulet,bless}.
The complementary nature of the prefill and decode phases makes them ideal for such concurrent execution.
Since LLM serving systems generally necessitate adherence to service-level objectives (SLOs) of predefined latency requirements \cite{distserve,splitwise,sarathi}, predictable and controllable execution time over the two phases is demanded.
However, current GPUs lack deterministic concurrent scheduling controls \cite{badmps,cudasched,libsmctrl2}, forcing users to carefully provision resources for kernels to achieve reliable overlap \cite{bless,orion,sdgrc}.
While modern GPUs provide compute resource partitioning through Nvidia's multi-process service (MPS) \cite{mps}, precise kernel management is still required to ensure effective resource sharing while meeting SLO requirements.
Figure \ref{fig:bg.sm_speedup} demonstrates that prefill scales near-linearly with SM count, while decode exhibits super-linear scaling.
This suggests potential throughput gains from concurrent execution with properly balanced SM allocation.

\textbf{Takeaway 1}. \textit{GPUs remain underutilized even during compute-intensive prefill. While co-locating prefill and decode saturates the resources, precise resource provisioning to orchestrate the two phases is demanded.}

\subsection{Biased Throughput-Latency Tradeoff}\label{sec:bg.biased}

\subsubsection{Chunked Prefill Workflow.}\label{sec:bg.chunk_workflow}

\begin{figure}[!b]
    \centering
    \vspace*{-1em}
    \includegraphics[width=\linewidth]{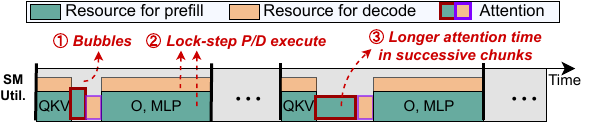}
    \includegraphics[width=\linewidth]{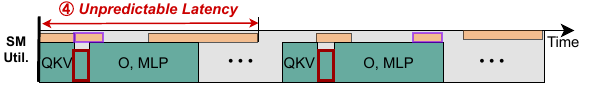}
    \vspace*{-2em}
    \caption{Kernel-level workflow of existing systems featuring chunked (top) and uncoordinated sharing (bottom). Both suffer from biased throughput-latency tradeoff.} 
    \label{fig:bg.compare}
\end{figure}

As shown in Figure \ref{fig:bg.compare}, chunked prefill \cite{sarathi} achieves low TPOT by leveraging a fixed token budget to concatenate the prefill and decode tokens into a \textit{hybrid batch}, executing in a lockstep fashion (\ding{183}).
Given a chunk size of $cs$, the hybrid batch is first filled with $ds$ active decode requests first, and allocates the remaining $cs-ds$ tokens to the prefill sequences.
Sequences $sl$ exceeding this residual capacity are split into chunks, leaving residual tokens processed in subsequent iterations.
Therefore, the prefill completion requires $N=\lceil sl / (cs - ds)\rceil$ iterations.
This forces $N\cdot(N+1)/2$ times KV cache reloads as each new chunk \textit{must} reprocess previous chunks' cached states.

Due to the lock-step execution of the hybrid batch, a smaller chunk size effectively decreases TPOT at the cost of increased TTFT and degraded system throughput \cite{sarathi}, while larger chunks exhibit the opposite effect.
Previous works \cite{sarathi,nanoflow} recognize these inherent tradeoffs and propose tuning chunk size based on workload's prefill-to-decode time ratio through manual tuning or automatic searching \cite{vidur,scoot}.

\subsubsection{Sub-optimal Hardware Utilization.}\label{sec:bg.chunkslow}

Despite the throughput-latency tradeoff has been extensively studied, we highlight that such a tradeoff is biased, and the resulting performance degradation remains overlooked.
\textbf{First}, as discussed in \cref{sec:bg.kernel}, chunked prefill typically uses suboptimal chunk sizes below GPU-saturating levels to prioritize low latency.
This produces severe wave quantization effects \cite{streamk}, creating GPU bubbles (Figure \ref{fig:bg.compare}-\ding{182}).
\textbf{Second}, redundant KV cache reloads required for long sequences significantly prolong attention computation time (\ding{184}), further reducing GPU utilization.
\textbf{Third}, these factors collectively inflate TTFT, triggering a cascading congestion effect in which queued requests stall while awaiting prefill completion, degrading overall system throughput.

\begin{figure}[!t]
    \centering
    \includegraphics[width=\linewidth]{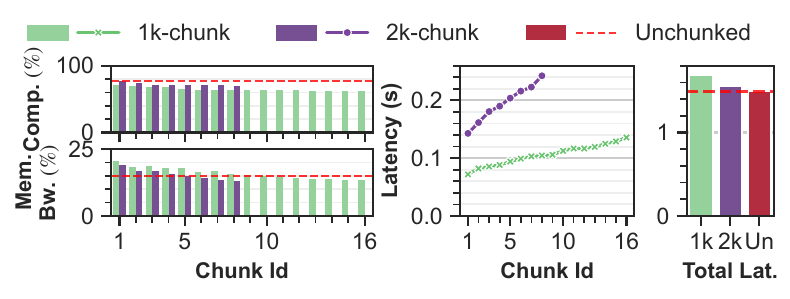}
    \begin{subfigure}[t]{.44\linewidth}
        \vspace{-2em}
        \caption{Hardware utilization.}
        \label{fig:bg.no_hybrid.hw}
    \end{subfigure}
    \hfill
    \begin{subfigure}[t]{.55\linewidth}
        \vspace{-2em}
        \caption{Per-chunk and total latency.}
        \label{fig:bg.no_hybrid.lat}
    \end{subfigure}
    \vspace*{-2em}
    \caption{Per-chunk GPU utilization and latency across varying chunk sizes, all exhibiting degraded throughput.}
    \vspace*{-1em}
    \label{fig:bg.no_hybrid}
\end{figure}

Figure \ref{fig:bg.no_hybrid} 
systematically quantify the performance degradation of chunked prefill of a 16k-token sequence prefill even \textit{without} hybrid batching decode requests.
For 1k chunk size, a progressive 10\% drop in compute efficiency (from 71\% to 61\%) across successive chunks is witnessed in Figure \ref{fig:bg.no_hybrid.hw}, which falls substantially below the 77\% achievable peak.
This under-utilization stems from redundant KV cache reloading in chunked attention, which also causes the final chunk's processing time to 1.9$\times$ than that of the initial chunk. 
Consequently, per-chunk latency scales linearly by chunk counts and increases total prefill latency by 1.13$\times$ compared to unchunked execution.
While a larger chunk size of 2k partially mitigates utilization drops from -18\% to -7\%, the average per-chunk latency increases by 1.86$\times$, significantly diminishing the TPOT improvements that motivated chunked prefill.
This fundamental tension between maintaining high hardware utilization and minimizing TPOT creates an intractable optimization in chunked prefill.

\begin{figure}[!b]
    \centering
    \vspace{-1em}
    \includegraphics[width=\linewidth]{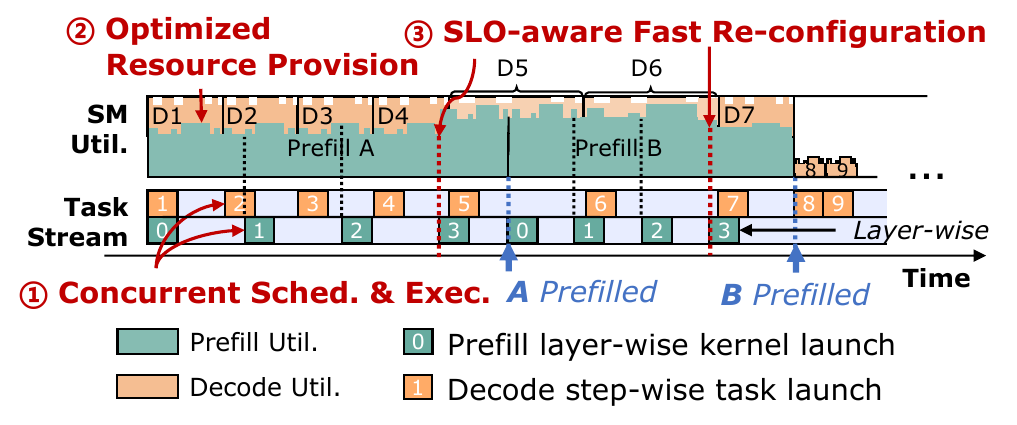}
    \caption{Dynamic spatial-temporal orchestration of concurrently executed prefill and decode tasks.}
    \label{fig:bg.design}
\end{figure}

\subsubsection{Existing Optimizations}
Despite recent optimizations of chunked prefill, the technique remains intrinsically limited. 
PodAttention \cite{podattn} fuses prefill and decode attention kernels to accelerate hybrid-batch attention, yet this operator-specific optimization does not mitigate the fundamental latency penalties imposed by chunking (\cref{sec:bg.chunkslow}).
Nanoflow \cite{nanoflow} designs a fixed-size pipeline that uses CUDA streams to overlap chunked attention and linear layers with careful inter-kernel synchronizations.
However, the latency of chunked attention grows by successive chunks, which eventually eliminates the chances of overlapping.
Apart from chunked prefill-based approaches, MuxServe \cite{muxserve} (Figure \ref{fig:bg.compare}) decouples prefill and decode phases into separate processes, leveraging MPS \cite{mps} with manually configured, fixed SM quotas to enable spatial sharing. 
This coarse-grained static partitioning leads to unpredictable latency despite saturating the GPU.
Therefore, neither solution fully resolves the throughput-latency tradeoff under varying serving demands.

\textbf{Takeaway-2}. \textit{Both chunked prefill and uncoordinated spatial sharing exhibit skewed throughput–latency tradeoffs. This urges for a fine-grained yet predictable execution mechanism for optimal LLM serving performance.}

\subsection{Opportunities and Challenges}

To address the inefficiencies discussed above, an inter-phase orchestration system is demanded to achieve optimal throughput-latency balance while maximizing GPU utilization, as illustrated in Figure \ref{fig:bg.design}.

\textbf{Concurrent Schedule and Execution.}
Co-executing prefill and decode eliminates redundant KV-cache reloads and fully utilizes GPU resources, but sustaining latency targets demands real-time request monitoring and kernel scheduling that respond to execution progress and system state.
A layer-level, SLO-aware scheduler must tightly control execution advance to avoid inter-phase resource competition.
While LLM online serving further requires low-overhead control plane and support for rapid reconfiguration, rendering these challenges non-trivial.

\textbf{Optimized Resource Provision.} 
By carefully partitioning available SM resources, memory-bound decode kernels may efficiently co-execute with compute-intensive prefill kernels.
However, LLM's multi-dimensional nature of input parameters and dynamic inter-kernel contention creates a complex optimization space.
This necessitates precise latency modeling across varying SM provisions and efficient exploration algorithms to identify optimal configurations.

\textbf{SLO-aware Rapid Re-configuration.}
In response to runtime workload dynamics, 
resource allocation between prefill and decode phases must be re-configured.
These frequent adjustments demand near-zero-overhead reconfiguration to maintain efficiency.
Although MPS \cite{mps} offers context APIs \cite{cuda} to adjust different SM quotas, it incurs significant memory overhead when applied to LLM serving scenarios due to dynamic inputs.
An even more lightweight approach is demanded to maintain for instantaneous switching between optimized resource partitions.

\section{Design}

\subsection{Overview}

\begin{figure}[!t]
    \centering
    \includegraphics[width=\linewidth]{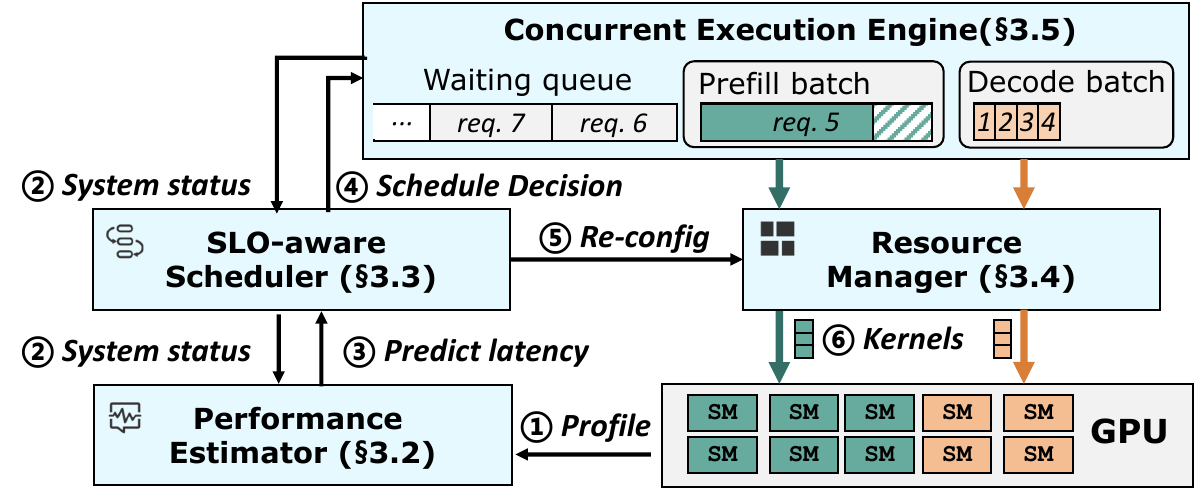}
    \caption{Workflow of \NAME{}. (Numbers: dataflow order.)}
    \label{fig:ds.overview}
    \vspace{-1em}
\end{figure}

Figure \ref{fig:ds.overview} illustrates \NAME{}'s workflow for concurrent prefill-decode execution with dynamic and fine-grained resource provisioning, with all components operating at microsecond-level overhead.
\NAME{} comprises four key components around the scheduling, resource partitioning and execution: performance estimator, SLO-aware task scheduler, resource manager, and concurrent execution engine.
The \textit{performance estimator} (\cref{sec:ds.estim}) \ding{182} first builds an analytical model for the served LLM, augmented with lightweight offline samples.
The model provides precise latency predictions across different configurations of co-executing prefill and decode batch sizes under varying resource allocations.
At runtime, the \textit{SLO-aware task scheduler} (\cref{sec:ds.sched}) acts as the central coordinator to enable GPU sharing between prefill and decode tasks via spatial-temporal scheduling.
During every layer-wise scheduling cycle, the scheduler \ding{183} proactively retrieves system status from the \textit{concurrent execution engine} (\cref{sec:ds.engine}), monitoring request progress and \ding{184} evaluating potential SLO violations by \textit{performance estimator}.
Observed statistics are also used to refine the performance estimator online.
The scheduler rapidly searches for \ding{185} an optimal resource configuration and scheduling decision that maximizes throughput while ensuring SLO compliance.
\textit{Computational resource manager} is \ding{186} triggered for lightning resource reconfiguration when necessary.
Finally, the prefill and decode kernels are \ding{187} launched concurrently on the provisioned SMs.
\NAME{} dynamically balancing the competing demands of TTFT and TPOT maintains high utilization with negligible overhead.

\subsection{Performance Estimator}\label{sec:ds.estim}

\subsubsection{Problem Formulation}\label{sec:ds.formulation}

For a given LLM and hardware, the latency of co-executed prefill and decode is determined by six factors, termed as Execution State ($ES$): prefill sequence length ($sl_i$), batch size ($pbs$) and number of allocated SMs ($pm$), alongside decode context length ($cl_i$), batch size ($dbs$) and SMs ($dm$).
Enumerating the millions of $ES$ combinations for profiling is infeasible.
Therefore, we build a lightweight analytical model with a minimal profile and runtime overhead.
First, we propose SM-scaling roofline model (SRM) to derive single kernel latency under partitioned SMs without interference. 
Second, memory-subsystem contention is quantified when concurrent kernels are isolated in distinct SMs.
Finally, the model is augmented with minimal sampled data for calibration.
At runtime, the model executes within microseconds and continuously refines with online statistics.

\subsubsection{SM-scaling Roofline Model (SRM)}

\begin{figure}[!t]
    \centering
    \includegraphics[width=\linewidth]{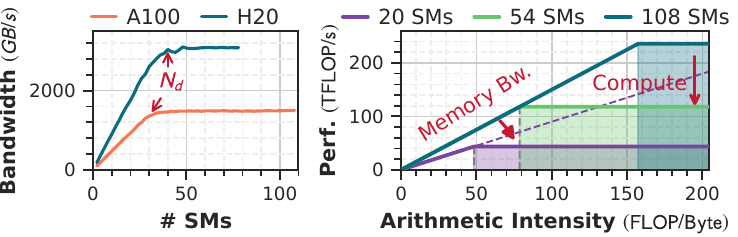}
    \begin{subfigure}[t]{.4\linewidth}
        \vspace{-1.5em}
        \caption{Achievable memory bandwidth by SM counts.}
        \label{fig:ds.bw.mem}
    \end{subfigure}
    \hfill
    \begin{subfigure}[t]{.5\linewidth}
        \vspace{-1.5em}
        \caption{SRM of A100 using measured peak performance.}
        \label{fig:ds.roofline.model}
    \end{subfigure}
    \vspace{-1em}
    \caption{Modeling peak performance by number of SMs.}
    \label{fig:ds.bw}
    \vspace{-1em}
\end{figure}

We examine the performance of compute, memory access, and network communication when only $N_p$ of the SMs are available. 
Theoretical compute performance scales linearly as $C_p=C_{peak}\cdot N_p/N$.
Memory and network bandwidth exhibit proportional scaling until reaching inflection points $N_d$ and $N_w$, 
where the SMs generate sufficient traffic to saturate the respective peak bandwidth $D_{peak}$ and $W_{peak}$.
Figure \ref{fig:ds.bw.mem} illustrates the throughput of a memory-copy kernel, showing the inflection points of Nvidia A100 and H20 GPUs.

Given a kernel with $flop_k$ operations and $mem_k$ bytes of memory transactions, we construct an SM-scaling roofline model (SRM) in Equation \ref{eq:ds.srm} to estimate the theoretical latency on $N_p$ SMs.
In the example of Figure \ref{fig:ds.roofline.model}, the memory bandwidth inflection point $N_d=30$.
When using 54 SMs, the attainable memory bandwidth remains at its peak, maintaining the original roofline slope while lowering the plateau.
For 20 SMs, both the slope and plateau decline.
\begin{equation}\label{eq:ds.srm}
    \begin{cases}
        T'_{k,p} = flop_k \cdot \min\left({flop_k}/{mem_k} \cdot D_p, C_p\right)^{-1} \\
        
        C_p = C_{peak} \cdot {N_p}/{N};\ D_p = D_{peak} \cdot \min(1, {N_p}/{N_d})\\
    \end{cases}
\end{equation}

\begin{figure}[t]
  \centering
  \includegraphics[width=\linewidth]{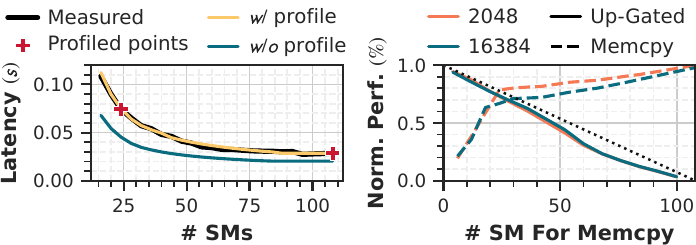}
  \begin{minipage}[t]{.45\linewidth}
    \centering
    \vspace{-2em}
    \caption{Calibrating estimated decode latency with only 2 samples.}
    \label{fig:ds.contention.srm}
  \end{minipage}\hfill
  \begin{minipage}[t]{.48\linewidth}
    \centering
    \vspace{-2em}
    \caption{Normalized performance of co-executed UG and memory copy.}
    \label{fig:ds.contention.memcpy}
  \end{minipage}
  \vspace{-1em}
\end{figure}

For each Execution State $ES$, we compute the arithmetic intensity of every LLM kernel with llm-viewer \cite{llmviewer}, apply SRM to obtain its baseline latency $T'_{k,p,ES}$ on Np SMs, and aggregate these values to yield the total LLM latency $T'_{p,ES}$.
Since practical execution rarely matches the roofline bound, Equation \ref{eq:ds.alpha} derives a scaling factor for calibrating SRM:
\begin{equation}\label{eq:ds.alpha}
    \alpha_{p,ES} = T^{\text{measured}}_{p,ES}/ T'_{p,ES}
\end{equation}
The factor is then extrapolated to unmeasured configurations, since the utilization pattern is near-linear between similar kernel inputs \cite{gpulet,habitat,baymax}.
As shown in Figure \ref{fig:ds.contention.srm}, only two samples are sufficient to model the decode latencies by varying SMs.
This demonstrates effective estimation aligned with kernel characteristics without extensive profiling.

\subsubsection{Modeling Contention}

When kernels execute on isolated SMs to prevent compute contention, memory subsystem and network contention persist.
We verify that kernel latency remains stable under such interference, even if underlying hardware scheduling and inter-kernel resource competition vary.
Extensive evaluation of 7360 concurrent $ES$ on A100 (serve Llama3.1-8B \cite{llama}) and 8$\times$H20 (serve Qwen3-32B \cite{qwen}), each repeated for 30 runs, observes that 95\% of latency measurements deviate less than $\pm$6.8\% from the mean of the respective repeated runs.
This tight distribution confirms the stability of kernel latency under varying prefill-decode interference patterns, which can be used as a reliable metric for contention modeling.

The worst-case memory subsystem interference can be quantified by co-executing a memory copy kernel on $N_p$ SMs and an up-gated layer (UG), which is a large-shaped GEMM in LLM, on the residual $N-N_p$ SMs.
This is because prefill and decode kernels typically exhibit lower resource utilization than both UG and memory copy.
Figure \ref{fig:ds.contention.memcpy} shows that UG exhibits marginal performance degradation (<8\%) when using more than 60\% SMs.
Therefore, the latency of prefill kernels can be reliably calibrated via Equation \ref{eq:ds.alpha} with minimal sample size.
Conversely, memory-copy throughput decreases as UG sequence lengths grow.
To model decode latencies, 
we measure the attainable bandwidth $D_{p,sl}$ when concurrently executed with $sl$-length prefill to update SRM.
Equation \ref{eq:ds.alpha} is then applied for further refinement, since the memory interference pattern scales linearly with kernel utilization characteristics \cite{gpulet,pythia}.

Inter-phase network contention is low since the traffic scales with either $sl$ or $dbs$, where $dbs$ is relatively small.
The memory subsystem impact of network transfers is inherently lower than memory copy, and prefill-to-decode interference is already accounted for in the end-to-end $D_{p,sl}$ calibration.

\subsubsection{Profiling and Online Calibration}

During offline profiling, \NAME{} records compute, memory, and network performance across varying SM counts in a single sweep to construct SRM, then measures prefill and decode latencies for a sparse set of $ES$ values as calibration.
Continuous online data collection for recalibration is straightforward.
Leveraging only Equation \ref{eq:ds.alpha} and linear extrapolation, the runtime overhead for update and prediction is negligible.

Table \ref{tab:des.accuracy} shows that calibrated SRM maintains high accuracy even for inputs beyond the sampled range while keeping profiling overhead low.
This precision is sufficient for reliable real-time scheduling optimization.

\renewcommand{\arraystretch}{0.9}
\begin{table}[!t]
    \centering
    \caption{Mean absolute relative error of SRM on 7360 data.}
    \vspace{-1em}
    \label{tab:des.accuracy}
    \begin{tabular}{c|c|c}
        \toprule
        Phase (Data Range) & \#Samples (Range) & MAPE \\
        \midrule
        & 18 ($sl\leq$16k) & 9.1\% \\
        
        \multirow{-2}{*}{\makecell{Prefill\\($sl\leq$32k)}} & 30 ($sl\leq$32k) & 5.3\% \\
        \midrule
        & 308 ($cl\leq$ 16k, $bs\leq$ 128) & 10.5\% \\
        
        \multirow{-2}{*}{\makecell{Decode\\($cl\leq$ 32k, $bs\leq$ 256)}} & 364 ($cl\leq$ 32k, $bs\leq$ 256) & 7.7\% \\
        \bottomrule
    \end{tabular}
    \vspace{-1em}
\end{table}
\renewcommand{\arraystretch}{1}

\subsection{SLO-aware Task Scheduler}\label{sec:ds.sched}

\subsubsection{Scheduling Workflow}\label{sec:ds.sched.workflow}

Each of the prefill and decode \textit{concurrent execution engine} (\cref{sec:ds.engine}) runs a scheduler autonomously.
At every step, the scheduler reads system status from the global metadata buffer (\cref{sec:ds.engine.kv}), forecasts latencies with \textit{performance estimator}, and reorders pending requests.
Upon detecting potential SLO violations, the scheduler greedily searches for the optimal configuration and invokes \textit{computational resource manager} (\cref{sec:ds.smpart}) to repartition SMs.

\begin{figure}[!b]
    \centering
    \vspace{-1.5em}
    \includegraphics[width=\linewidth]{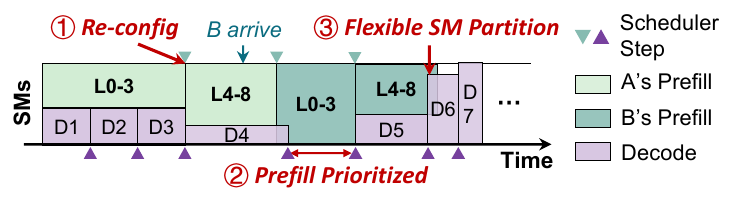}
    \vspace{-2em}
    \caption{Asynchronous scheduler launches layer-wise prefill kernels and step-wise decode graphs, monitoring latencies and dynamically reallocating SMs.}
    \label{fig:ds.scheduler}
\end{figure}

As illustrated in Figure \ref{fig:ds.scheduler}, for prefill scheduler, a fixed number of layers is launched per step, and synchronizes for CPU (teal triangle) to make subsequent scheduling decisions.
This enables fine-grained control over prefill progress temporally for rapid adaptation to system fluctuations.
Conversely, decode scheduler issue kernels as a single CUDA Graph \cite{cuda} to eliminate the launch overhead of small kernels.

The primary scheduling objective is to prioritize prefill while respecting decode SLOs, since shorter prefill latency enlarges decode batch size and raises system throughput \cite{sarathi}.
During concurrent operation, the decode phase is provisioned with the minimum SM counts that satisfy SLO.
As the final prefill layer approaches completion, additional SMs are allocated to decode phase to facilitate a smooth transition between co-running prefill/decode and decode-only.
During reconfiguration, \NAME{} eliminates inter-phase synchronization by partially sharing SMs between phases rather than idling unused resources. 
Although a perfect non-overlapping allocation is challenging, layer-wise prefill scheduling confines such compute resource interference to minimal, predictable regions.

\begin{algorithm}[!b]
\caption{SLO-aware Scheduling for Prefill}
\label{alg:scheduler}
\DontPrintSemicolon 
\KwIn{System state: $S$,\quad TTFT/TPOT SLO: $\Gamma_p$/$\Gamma_d$}
\KwOut{Next requests \& layers to run: $next\_tasks$}
\SetKwFunction{FMain}{Schedule} 
\SetKwFunction{FReduce}{ReduceDecodeSM} 
\SetKwProg{Fn}{Function}{:}{end} 

\Fn{\FMain{$S$}}{
    
    $ttft \gets \Call{EstimateLatency}{S}$\;
    $\Call{WriteGlobalBuffer}{ttft}$\;
    $tpot \gets \Call{ReadGlobalBuffer}{ }$\;
    
    $\Call{SortByLeastEstimLatency}{Q}$\; 
    \uIf{$P \neq \emptyset$}{
        $satisfy \gets \text{req.ttft} \leq \Gamma_p, \text{req}\in P$\;
        $next\_tasks \gets P $\;
    }
    \Else{
        $satisfy \gets \Call{P90}{ttft} \leq \Gamma_p $\;
        $next\_tasks \gets \emptyset$\; 
        \While{\Call{ArithInten}{next\_tasks, S.ES} < peak}{
            $next\_tasks.\text{append}(Q.\text{pop()})$\; 
        }
    }

    \uIf{$\textbf{not } \text{satisfy} \textbf{ and } \Call{P90}{tpot} > \Gamma_d$}{
        {$\Call{SetBalancedSM}{S, ttft, tpot}$}\;
    }
    \ElseIf{$\Call{P90}{tpot} \leq \Gamma_d$}{
        {$\Call{ReduceDecodeSM}{S, ttft, tpot}$}\;
    }
    \ElseIf{$\Call{P90}{tpot} > \Gamma_d$}{
        {$\Call{ReducePrefillSM}{S, ttft, tpot}$}\;
    }

    \Return{$\text{next\_tasks}, , L_{exe}+L_{step}$}
}
\end{algorithm}

\subsubsection{Request Scheduling and Resource Provisioning}

To facilitate SLO-aware scheduling, \NAME{} monitors the system state defined as $S=(ES,PS,RS)$, where $ES$ denotes the execution state (\cref{sec:ds.formulation}), $PS$ the prefill progress, and $RS$ the per-request latency metrics. 
Prefill state $PS$ comprises the queuing-request set $Q$, the in-flight request set $P$, and the executed layer count $L_{exe}$.
For every request $i$, arrival time $a_i$ and decode-start time $d_i$ are recorded.
At time $t$, the prefill scheduler estimates GPU execution times for all requests in $P \cup Q$ under current $ES$, while the decode scheduler predicts next-step latency and updates the corresponding TPOT.
These estimations are then written to the global metadata buffer, allowing each scheduler perceive the holistic state.

Algorithm \ref{alg:scheduler} outlines the prefill scheduler, and the decode scheduler follows analogously.
The algorithm continuously monitors execution progress and updates latency estimates (lines 2-4).
Requests in the waiting queue are reordered by ascending predicted latency (line 5) when the reordering does not violate TTFT SLOs for pending requests.
This reduces average TTFT without starvation.
To start a new prefill step (line 10-13), requests are batched until reaching arithmetic intensity limits under the current execution state.
When provisioning resources, prefill phase is prioritized and provisioned with more SMs unless TPOT is compromised (line 15).
In extreme cases of high request load, the decode phase would be temporarily suspended (Figure \ref{fig:ds.scheduler}-\ding{193}) if TPOT SLO is still met.
When both SLOs cannot simultaneously be satisfied, indicating the system is beyond maximum capacity, a balanced SM ratio is enforced to limit excessive latency in either phase (line 14).

\subsection{Computational Resource Manager}\label{sec:ds.smpart}

While MPS \cite{mps} with CUDA Green Context \cite{greenctx} supports SM partitioning, its memory overhead exceeds 700MB for only 4 static policies in LLM serving \cite{drift}, rendering it impractical for \NAME{}'s fine-grained control.
For flexible low-overhead SM provisioning, we employ a library of SM masking techniques \cite{libsmctrl,libsmctrl2}.
The \texttt{libsmctrl\_set\_stream\_mask} API sets the metadata of a CUDA stream \cite{cuda} to restrict all kernels launched within it to execute on a specified SM subset.
Section \ref{sec:eval.overhead} validates that this approach adds only microsecond-level runtime overhead and zero extra memory. 
Similar interfaces exist on other platforms, exemplified by AMD's \texttt{hipExtStreamCreateWithCUMask} \cite{hipstream}.

\NAME{} creates a CUDA stream in each \textit{concurrent execution engine} for prefill and decode.
Whenever the scheduler issues a repartitioning command, the system immediately invokes the \texttt{libsmctrl} API to reconfigure the corresponding stream, thereby restricting all subsequent kernels to the newly provisioned SMs.
This instantaneous configuration (Figure \ref{fig:ds.scheduler}-\ding{192}\ding{194}), supports rapid adaptation to dynamic system states with flexible SM allocation.

\subsection{Concurrent Execution Engine}\label{sec:ds.engine}

\begin{figure}[!t]
    \centering
    \includegraphics[width=\linewidth]{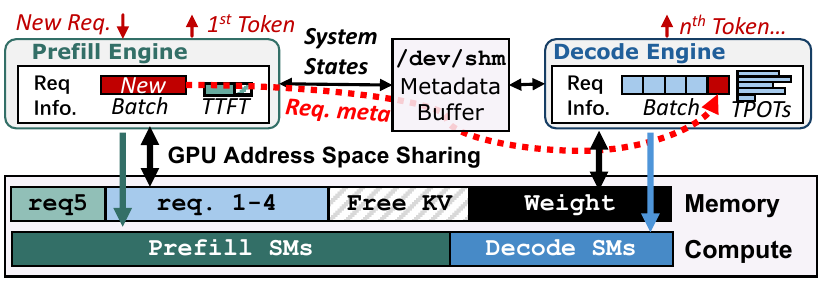}
    \vspace*{-2em}
    \caption{Concurrent execution engine with shared KV cache/weight GPU address spaces, exchanging system states and request metadata via OS-managed shared memory.}
    \label{fig:ds.engine}
    \vspace*{-1em}
\end{figure}

\subsubsection{Engine Architecture}\label{sec:ds.engine.kv}

Figure \ref{fig:ds.engine} illustrates the concurrent execution engines for prefill and decode, each residing in a separate process and driven by the corresponding scheduler.
Both engines share a CPU buffer and unified GPU memory pool.
The CPU buffer is implemented as OS-managed shared memory, stores global system states and employs compact control bits to indicate data availability, enabling low-latency status exchange.
For GPU memory management, \NAME{} employs a dedicated initialization process that allocates model weights and KV cache \cite{vllm} prior to engine launch.
The resulting memory region is shared between engines via \texttt{cudaIpcGet/OpenMemHandle} API \cite{cuda} with no adverse effects as documented.
Equivalent facilities, such as AMD’s \texttt{hipIpcOpenMemHandle} \cite{hipstream}, allow the same design to be deployed on other hardware.
Since the address space and metadata are shared, \NAME{} is fully compatible with existing KV cache and prefix cache optimizations. 
An atomic lock serializes allocation and deallocation transactions that may be issued concurrently by both engines, ensuring correctness with minimal performance impact.

\subsubsection{Execution Workflow}
The prefill engine's execution begins with receiving requests and follows the scheduling workflow in \cref{sec:ds.sched}, retrieving decode-side states from the buffer, such as batch size and TPOTs.
Once prefill completes, the request migrates to decode without KV-cache transfer, and only metadata is asynchronously sent to the decode engine via ZeroMQ \cite{zeromq}, enabling microsecond overhead.
The decode engine receives newly prefilled requests and merges them into the current running batch.
Output tokens are directly forwarded to the frontend server, eliminating any CPU involvement by the prefill engine.
\NAME{}'s control plane works independently while proactively communicating through the buffer, unblocking both CPU and GPU execution.
This decentralized architecture allows concurrent kernel submissions while eliminating the need for frequent synchronization compared with a centralized architecture.
We implement \NAME{} on top of SGLang \cite{sglang} v0.4.6 and PyTorch 2.6.0 with 4100 lines of Python code, and integrate a modified \texttt{libsmctrl} \cite{libsmctrl} library to optimize GPU resource allocation within the serving engine.
The prefill and decode engines are implemented as SGLang's workers, with MPS \cite{mps} enabled for spatial sharing.

\section{Experimental Evaluation}

\subsection{Methodology}\label{sec:eval.metho}

\textbf{Evaluated Schemes.} We compare \NAME{} against several state-of-the-art systems of chunked prefill-based optimization: vLLM \cite{vllm} v0.8.5 (1024-chunk), SGLang \cite{sglang} v0.4.6 (1024/2048-chunk), and Nanoflow \cite{nanoflow} (1024-chunk).

\begin{figure}[!b]
    \centering
    \vspace{-1em}
    \includegraphics[width=\linewidth]{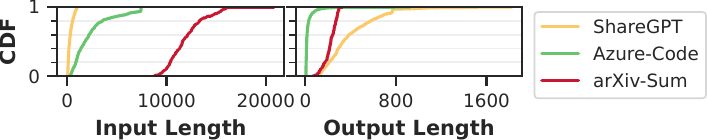}
    \vspace{-2em}
    \caption{Workload input/output length CDF.}
    
    \label{fig:eval.dataset}
\end{figure}

\begin{figure*}[!t]
    \centering
    \includegraphics[width=\linewidth]{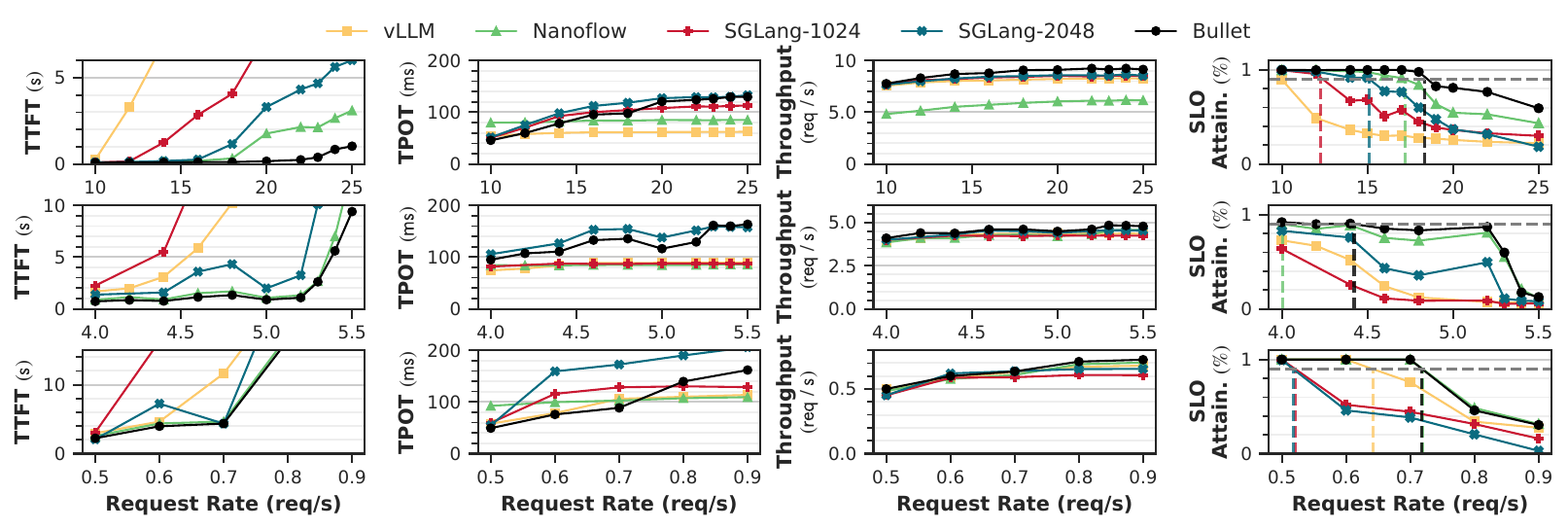}
    \hspace*{1em}
    \begin{subfigure}[t]{.24\linewidth}
        \vspace*{-2em}
        \caption{Average TTFT.}
        \label{fig:eval.e2e.ttft}
    \end{subfigure}
    \hfill
    \begin{subfigure}[t]{.23\linewidth}
        \vspace{-2em}
        \caption{Average TPOT.}
        \label{fig:eval.e2e.tpot}
    \end{subfigure}
    \hfill
    \begin{subfigure}[t]{.25\linewidth}
        \vspace{-2em}
        \caption{Request throughput.}
        \label{fig:eval.e2e.thrput}
    \end{subfigure}
    \hfill
    \begin{subfigure}[t]{.24\linewidth}
        \vspace{-2em}
        \caption{P90 SLO attainment.}
        \label{fig:eval.e2e.slo}
        \vspace*{-2em}
    \end{subfigure}
    \vspace{-2em}
    \caption{Performance comparison of ShareGPT (top row), Azure-Code (middle) and arXiv-Summary (bottom) datasets, evaluating average latency, throughput and SLO attainment rate of different LLM serving systems. \NAME{} achieves the highest throughput and SLO compliance in all workloads, attributed to remarkably decreased TTFT with modest TPOT increments.}
    \label{fig:eval.e2e}
    \vspace{-.5em}
\end{figure*}

\textbf{Workloads}
Three representative datasets are used for evaluation, whose request arrivals follow a Poisson process, as widely adopted in prior works \cite{vllm,sarathi,nanoflow}.
Figure \ref{fig:eval.dataset} details the cumulative distribution function (CDF) of the datasets' sequence length.
The ShareGPT \cite{sharegpt} contains real-world conversational data, Azure-Code \cite{splitwise} is production code completion trace released by Azure, and arXiv-Summary \cite{alpaca1} is long-context summarization.

\textbf{Models and Platforms.}
Experiments are conducted on three servers in Table \ref{tab:eval.server} with CUDA 12.4.
We evalute dense model Llama3.1-8B/70B \cite{llama} and MoE model Qwen3-235B-A22B-FP8 \cite{qwen}.
All multi-GPU setups use tensor parallelism.

\renewcommand{\arraystretch}{0.8}
\begin{table}[!b]
    \centering
    \vspace{-1em}
    \caption{Server configurations for experiments.}
    \vspace{-1em}
    \label{tab:eval.server}
    \begin{tabular}{ccccc}
        \toprule
        GPUs & Network Bw. & \#SMs/GPU \\
        \midrule
        8$\times$A100-80GB & 20 GB/s & 108 \\
        8$\times$H100 & 600 GB/s & 132 \\
        8$\times$H20 & 400 GB/s & 78 &  \\
         \bottomrule
    \end{tabular}
    
\end{table}
\renewcommand{\arraystretch}{1.0}

\renewcommand{\arraystretch}{0.8}
\begin{table}[!b]
    \centering
    \vspace{-1em}
    \caption{Workload latency requirements.}
    \vspace*{-1em}
    \begin{tabular}{c|ccc}
        \toprule
        & ShareGPT & Azure-Code & arXiv-Sum \\
        \midrule
        norm. TTFT & 3.0 ms & 1.5 ms & 1.5 ms \\
        TPOT & 150 ms & 200 ms & 175 ms \\
        \bottomrule
    \end{tabular}
    \label{tab:eval.slo}
    
\end{table}
\renewcommand{\arraystretch}{1.0}

\textbf{Metrics.}
We report TTFT, TPOT and throughput.
For SLO compliance measurement, we define as prefill and decode latency satisfying both the constraints listed in Table \ref{tab:eval.slo}, aligned with previous works \cite{nanoflow,loongserve,distserve}.
Due to the variation in sequence length, we use \textit{normalized input/generation latency} established in previous works \cite{loongserve,nanoflow} for SLO measurement.
We collect hardware metrics using Nsight Systems \cite{nsys}.

\subsection{End-to-End Performance}

\subsubsection{Single GPU Performance}

Figure \ref{fig:eval.e2e} evaluate \NAME{} against four baselines across three real-world workloads. 
\NAME{} demonstrates consistent throughput improvements, achieving 1.09$\times$ average and up to 1.20$\times$ higher throughput compared to SGLang-1024.
Overall, \NAME{} maintains superior prefill latency enhancement (13.5$\times$) with acceptable TPOT (0.94$\times$) in average through dynamic SM allocation, translating to 1.86$\times$ end-to-end speedup compared with SGLang-1024.
In contrast, all the chunked prefill-based systems exhibit unacceptable TTFT degradation even in a low request rate.
For example, on the ShareGPT workload, with request rate of 20req/s, the prefill latency ranges from 1.8s (Nanoflow) to 14.9s (vLLM).
The key reason is that chunk prefill limits the system's capability to process the prefill phase (\cref{sec:bg.chunkslow}) and creates a cascading congestion for pending requests.

\subsubsection{SLO Attainment}

Figure \ref{fig:eval.e2e.slo} presents P90 SLO attainment rate.
By eliminating these bottlenecks through intelligent SM allocation and adaptive scheduling, \NAME{} achieves mean TTFT of 0.16s and P90 tail latency of 0.31s, which is 54.9$\times$ and 78.5$\times$ better to SGLang-1024.
The lower prefill latency remarkably translates to larger decode batches that boost overall throughput and SLO compliance (1.49$\times$).
\NAME{} effectively balances the throughput-latency tradeoff that is skewed in chunked prefill systems. 
While SGLang-2048 demonstrates a 3.20$\times$ TTFT improvement and 1.18$\times$ higher throughput than SGLang-1024, the 2048 chunk suffers from 0.78$\times$ worse TPOT on average, confirming the imbalanced tradeoff inherent in chunked prefill.
\NAME{} breaks this paradigm by simultaneously achieving both lower TTFT (4.2$\times$) and better TPOT (1.20$\times$) than SGLang-2048 through exploiting GPU utilization via concurrent prefill-decode execution.
While Nanoflow achieves 2.37$\times$ longer TTFT and 0.86$\times$ shorter TPOT than \NAME{} via static kernel overlapping pipeline, it still suffers from the inherent limitations of chunked prefill that inflate TTFT tail latency.
This results in 5.2\% lower SLO compliance and 20.0\% lower throughput compared to \NAME{}.

\subsubsection{Multi-GPU Performance}

\begin{figure}[!b]
    \centering
    \includegraphics[width=\linewidth]{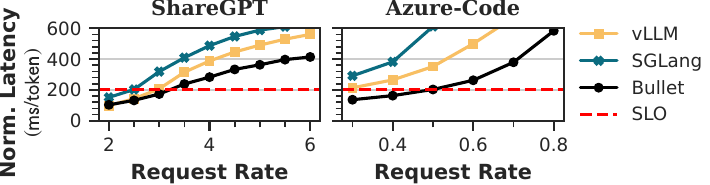}
    \vspace{-2em}
    \caption{Mean normalized generation latency of Llama3.1-70B served on 8$\times$A100. \NAME{} sustains higher request rates while remaining within the SLO constraint.}
    \label{fig:eval.a800}
\end{figure}

Figure \ref{fig:eval.a800} depicts the generation latency of \NAME{} when Llama3.1-70B is deployed across eight A100 GPUs, with vLLM and SGLang using a chunk size of 2048.
On the ShareGPT workload at 3.0 req/s, \NAME{} achieves a mean latency of 173 ms/token (223 ms/token for P99), whereas vLLM and SGLang attain 207 ms/token and 319 ms/token, respectively.
For the Azure-Code dataset, where input lengths are significantly longer, \NAME{}'s advantages is more pronounced, sustaining 203 ms/token at 0.5 req/s. 
In contrast, vLLM and SGLang already exhibit 213 ms/token and 291 ms/token at a lower rate of 0.3 req/s.

On H100 GPUs, the elevated memory bandwidth mitigates chunked-prefill overhead, yet \NAME{} continues outperforms SGLang and vLLM.
\NAME{}'s concurrent handling of prefill and decode tasks leverages potential overlaps between computation, communication and memory operations with fine-grained SM control between phases.
We also demonstrate \NAME{}'s generalization to other LLMs.
For the highly sparse, FP8-quantized MoE Qwen3-235B-A22B, chunked prefill's inherent lock-step batching impedes both prefill and decode speeds.
However, \NAME{} orchestrates the two phases effectively.
At 4.0 req/s, \NAME{} delivers 110 ms/token, with 1.4s TTFT and 45ms TPOT—respectively, which is 17.7$\times$, 16.2$\times$ and 4.2$\times$ lower than vLLM.

\begin{figure}[!t]
    \centering
    \includegraphics[width=\linewidth]{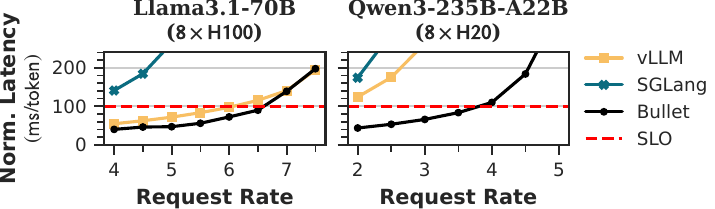}
    \caption{Mean normalized generation latency of various models across different hardware serving Azure-Code.}
    \label{fig:eval.hopper}
    \vspace{-1em}
\end{figure}

\begin{figure}[!b]
    \centering
    \vspace{-1em}
    \begin{subfigure}[t]{\linewidth}
        \includegraphics[width=\linewidth]{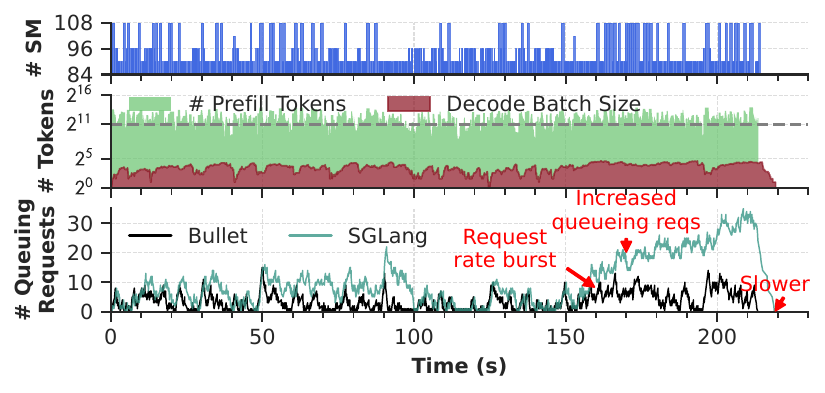}
        \vspace{-2em}
        \caption{SM allocation for prefill changes dynamically according to system load (top). Concurrently processing tokens/batch size in \NAME{} (middle). Number of requests waiting for prefill (bottom). \NAME{} adaptively paritition SMs to avoids request congestion on burst.}
        \label{fig:eval.break.timeline_my}
    \end{subfigure}
    \begin{subfigure}[t]{\linewidth}
        \includegraphics[width=\linewidth]{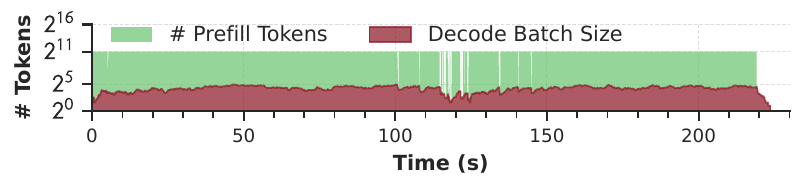}
        \vspace{-2em}
        \caption{SGLang's hybrid batch status. Decode requests occupy chunk budget, degrade prefill efficiency, incurring severe queuing time.}
        \label{fig:eval.break.sglang}
        \vspace*{-1em}
    \end{subfigure}
    \caption{System serving status of the Azure-Code workload in timeline view (request rate: 5.0 req/s). }
    
    \label{fig:eval.break.timelines}
\end{figure}

\subsection{Performance Breakdown}

\subsubsection{Dynamic SM Partition}

We break down the effectiveness of dynamic resource provisioning in Figure \ref{fig:eval.break.timelines} by timeline with the Azure-Code workload (5.0 req/s, Llama3.1-8B).
The top row in Figure \ref{fig:eval.break.timeline_my} demonstrates the number of SMs provisioned for the prefill phase, with each bar showing the SM count and duration.
On request rate bursts (spikes in Figure \ref{fig:eval.break.timelines}-bottom), \NAME{} adaptively sets the number of prefill SMs to use full GPU, and may temporarily delay decode requests.
This enables the rapid response for queuing requests, avoiding excessive waiting time.
After processing the pending requests, \NAME{} quickly reconfigurates the resources to a balance point for both phases.
Instead, chunked prefill-based systems are not able to optimize such a scenario, since redundant KV cache access inflates the prefill speed.
Moreover, decode batch shares the token budget with prefill tokens (Figure \ref{fig:eval.break.sglang}), enforcing more iterations to complete a prefill, which further degrades performance.
These factors jointly result in 4.17$\times$ longer queuing delay for SGLang-2048.
\NAME{} exempts from the token budget (Figure \ref{fig:eval.break.timelines}-middle) and orchestrates the two phases with fine-grained resource control to saturate GPU, significantly decreasing both TTFT and TPOT by 9.15$\times$ and 1.33$\times$, respectively.

\subsubsection{GPU Utilization}

\begin{figure}[!t]
    \centering
    \includegraphics[width=\linewidth]{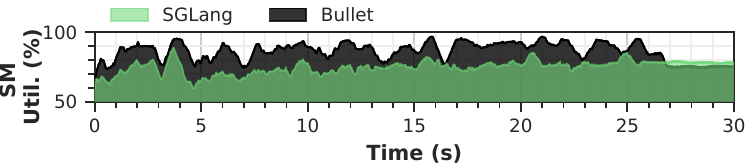}
    \caption{SM utilization of a timeslice in ShareGPT workload (20 req/s, Llama3.1-8B).}
    \label{fig:eval.utilization}
    \vspace{-1em}
\end{figure}

We measure hardware utilization with Nsight Systems and present it in Figure \ref{fig:eval.utilization}.
Between 0 s and 27 s, when the system concurrently handles prefill and decode requests, \NAME{} sustains an average of 86.2\% active SM cycles, which is 11.2\% higher than SGLang. 
Correspondingly, utilization of Tensor Cores rises by 11.8\%, while memory-bandwidth utilization increases by 19.3\%. 
These utilization gains directly reduce the mean end-to-end request latency from 25.8 s to 21.4 s.

\subsubsection{Overheads}\label{sec:eval.overhead}

Table \ref{tab:eval.break.overhead} quantifies the CPU overhead of \NAME{}'s key components.
Sending and receiving metadata only necessitates serialization and deserialization of Python objects, therefore achieving mean latency of 0.21$ms$.
The performance prediction module incurs merely 10.2$\mu s$ overhead by simply invoking the analytical model. 
For GPU resource management, SM masking enables instantaneous reconfiguration with
negligible runtime overhead. 
These designs jointly ensure \NAME{}'s control plane adds minimal latency while enabling fine-grained resource optimization.

\subsection{Sensitivity Studies}\label{sec:eval.sensi}

\begin{figure}[!b]
    \centering
    \vspace{-1em}
    \includegraphics[width=\linewidth]{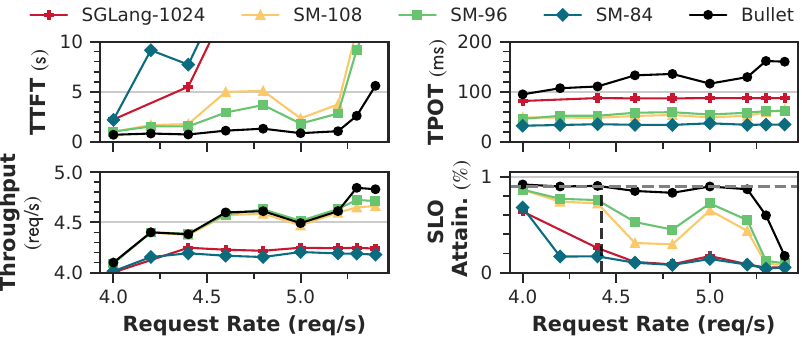}
    \caption{Sensitivity study on fixed SM counts for prefill. Static setups cause latency-throughput imbalance: more SMs cut TTFT but harm TPOT and goodput. \NAME{}'s dynamic SM tuning optimizes both.}
    \label{fig:eval.sensi.big}
\end{figure}

To study the effect of dynamic resource provisioning, we run the workloads under fixed SM configurations for prefill and allow decode to use all SMs.
The results are reported in Figure \ref{fig:eval.sensi.big}.
The SM-108 configuration (no partitioning) demonstrates severe imbalance in the Azure-Code workload.
While achieving low TPOT, SM-108 suffers 1.20$\times$ higher TTFT in average and 1.19$\times$ worse P90 tail latency compared to \NAME{}, ultimately reducing throughput and SLO attainment by 13\%. 
Smaller static partitions, like 84 SMs, prove even more problematic, exacerbating latency imbalance with 1.78$\times$ worse TTFT even than chunked prefill baselines and reducing throughput by 5.9\%.
Therefore, there is no optimal fixed SM allocation, as smaller partitions improve TPOT but degrade TTFT and introduce tail latency violations, and vice versa.
These results confirm \NAME{}'s dynamic approach, which adaptively orchestrates resources to simultaneously maintain balanced latency targets and maximize throughput.

\renewcommand{\arraystretch}{0.9}
\begin{table}[!t]
    \centering
    \caption{Overheads in \NAME{}.}
    \vspace{-1em}
    \begin{tabular}{l|cccc}
        \toprule
        & \textbf{Mean} & \textbf{Std.} & \textbf{P90} & \textbf{P99} \\
        \midrule
        
        \textbf{Metadata Send/Recv} (ms) & 0.21 & 0.44 & 0.89 & 1.54 \\
        \textbf{Performance Predict} ($\mu$s) & 10.2 & 5.1 & 24.5 & 25.8 \\
        \textbf{Resource Re-config} ($\mu$s) & 4.1 & 0.79 & 4.2 & 5.9 \\
        \bottomrule
    \end{tabular}
    \vspace{-1em}
    \label{tab:eval.break.overhead}
\end{table}
\renewcommand{\arraystretch}{0.9}

\subsection{Ablation Studies}

\begin{figure}[!t]
    \centering
    
    \begin{subfigure}[t]{\linewidth}
        \includegraphics[width=\linewidth]{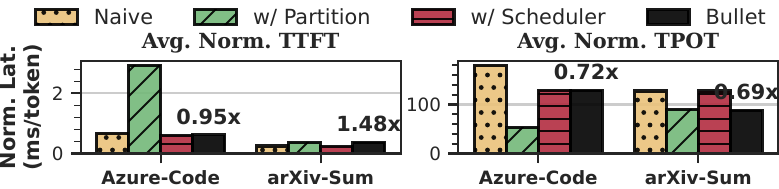}
    \end{subfigure}
    \vspace{-2em}
    \caption{Ablation study of \NAME{}.} 
    \label{fig:eval.abla.big}
    \vspace{-1em}
\end{figure}

Figure \ref{fig:eval.abla.big} analyzes the contributions of different components in \NAME{} by isolating part of the design. Naive: concurrent execution without resource provisioning or scheduling. \textit{w/}Partition: adds resource provision only. \textit{w/}Scheduler: Incorporates only request reordering and delayed decode.
The Naive design exhibits the expected latency imbalance (\cref{sec:eval.sensi}), high TPOT and low TTFT attributed to unpartitioned resource contention.
The \textit{w}/Partition variant improves TPOT for Azure-Code but suffers unacceptable TTFT degradation from its inability to reorder pending requests. 
Conversely, \textit{w}/Scheduler maintains comparable latency while reducing Azure-Code TPOT through contention alleviation.
Only the complete \NAME{} design, combining both partitioning and scheduling, achieves balanced latency across all workloads.

\subsection{Discussions}

While \NAME{} shows promising improvements, we identify several operational boundaries.
First, on low-compute GPUs running dense models, the limited SM budget obliges the decode phase to claim a larger ratio of SMs to saturate memory bandwidth (Figure \ref{fig:ds.bw.mem}), slightly tempering concurrent-execution benefits.
However, this caveat is immaterial for MoE models, whose lighter compute footprint preserves the full advantage (Figure \ref{fig:eval.hopper}).
Second, \NAME{}'s SM isolation policy could be relaxed to partial sharing, yet such exploration is orthogonal to the paper’s core contribution of fine-grained spatial–temporal orchestration and can be retrofitted without architectural changes.
Third, for specialized architectures like DeepSeek-MLA \cite{dsv3} where disaggregation proves fundamentally advantageous, \NAME{}'s co-located approach may not match dedicated solutions' peak performance. 
Extending the latency model to cover new attention variants or LoRA adapters \cite{lora} is straightforward and left to future work.
\NAME{}'s core innovations maintain broad applicability across standard LLM architectures and typical workload distributions while extensible to emerging models.

\section{Related Works}

\textbf{LLM Serving Systems.}
System-level optimizations for LLM serving have been extensively explored, which can be categorized as prefill-decode co-location and disaggregation.
Co-located techniques represented by chunked prefill \cite{sarathi,nanoflow} improve batching efficiency but maintain lockstep prefill and decode phase execution that underutilizes hardware. 
Disaggregated systems \cite{distserve,splitwise,cacheattn,mooncake} eliminate inter-phase interference but incur costly KV cache migration and load imbalance among prefill-decode instances. 
\NAME{} uniquely enables intra-GPU prefill-decode execution, maximizing utilization without overheads while meeting SLOs.
Recent intra-GPU sharing proposals for LLM inference remain fundamentally constrained. 
Nanoflow \cite{nanoflow} exploits kernel overlap atop chunked prefill. 
However, the approach’s intrinsic limitations erode gains as sequence lengths increase.
MuxServe \cite{muxserve} and Semi-PD \cite{semipd} rely on static SM allocations, requiring costly engine relaunches for reconfiguration.
Drift \cite{drift} explores lock-step execution between prefill blocks and decode in gangs using predefined SM partitions.
These coarse designs lack the agility required by dynamic workloads.
\NAME{} addresses these shortcomings through adaptive fine-grained resource management, enabling optimal SM utilization across diverse workloads.

\textbf{Concurrent Kernel Execution.}
Co-executing kernels with complementary resource demands has been proven to improve GPU utilization and performance.
Spatial-temporal multiplexing techniques \cite{krisp,gslice,gshare} typically classify kernels by computational characteristics and rely on CUDA streams \cite{cuda} or MPS \cite{mps} for kernel submission.
However, these approaches treat the hardware scheduler as a blackbox, providing limited control \cite{badmps,cudasched} over actual execution.
While several systems use precise kernel timing and resource provisioning \cite{bless,gpulet,orion} to enforce predictable scheduling, they focus on traditional DNN workloads and fail to address LLM-specific requirements, including prefill-decode phase coordination and KV cache management.
\NAME{} distinguishes itself from previous works by fine-grained orchestration of LLM's phases and layer-level precise resource management.

\textbf{Kernel Scheduling.}
Effective scheduling is essential for optimizing concurrent kernel execution.
Latency model-based solutions \cite{bless,gpulet,orion} schedule kernels within latency constraints but rely on excessive offline profiling, which is unsuitable for the dynamic workloads in LLM serving.
Although estimating standalone LLM inference latency is well-established \cite{vidur,llmviewer,loongserve}, the complex interactions between concurrent prefill and decode phases remain unstudied.
Existing solutions either rely on simplified models \cite{semipd} or assume contention-free execution \cite{drift}.
\NAME{} addresses this gap by developing a profile-augmented analytical model for performance estimation and refining kernel scheduling mechanisms for LLM serving.

\vspace{-1em}

\section{Conclusion}
This paper proposes \NAME{} to optimize LLM serving through precise GPU resource provisioning.
By carefully orchestrating spatial-temporal execution between prefill and decode phases, \NAME{} effectively saturates wasted GPU resources.
At runtime, \NAME{} employs real-time task progress estimation to optimize scheduling decisions and resource allocation, enabling precise and fine-grained control over layer-level resource quotas.
Experimental evaluation confirms consistent latency compliance and substantial throughput gains.

\balance

\bibliographystyle{ACM-Reference-Format}
\bibliography{ref}


\end{document}